# On the Inverse Of General Cyclic Heptadiagonal and Anti-Heptadiagonal Matrices

A.A. KARAWIA[1]
Computer Science Unit, Deanship of Educational Services,
Qassim University, Buraidah 51452, Saudi Arabia.
kraoieh@qu.edu.sa

## ABSTRACT

In the current work, the author present a symbolic algorithm for finding the determinant of any general nonsingular cyclic heptadiagonal matrices and inverse of anti-cyclic heptadiagonal matrices. The algorithms are mainly based on the work presented in [A. A. KARAWIA, A New Algorithm for Inverting General Cyclic Heptadiagonal Matrices Recursively, arXiv:1011.2306v1 [cs.SC]]. The symbolic algorithms are suited for implementation using Computer Algebra Systems (CAS) such as MATLAB, MAPLE and MATHEMATICA. An illustrative example is given.

## 1. INTRODUCTION

The n × n general periodic heptadiagonal matrices are takes the form:

$$H = \begin{bmatrix} d_1 & a_1 & A_1 & C_1 & 0 & 0 & \cdots & & B_1 & b_1 \\ b_2 & d_2 & a_2 & A_2 & C_2 & \ddots & \ddots & \cdots & & B_2 \\ B_3 & b_3 & d_3 & a_3 & A_3 & \ddots & \ddots & \ddots & \cdots & \\ D_4 & B_4 & b_4 & d_4 & a_4 & \ddots & \ddots & \ddots & \ddots & \\ 0 & \ddots & \ddots & \ddots & \ddots & \ddots & \ddots & \ddots & \ddots & \vdots \\ \vdots & \ddots & \ddots & \ddots & \ddots & \ddots & \ddots & \ddots & \ddots & \\ 0 & \ddots & \ddots & \ddots & \ddots & \ddots & \ldots & a_{n-3} & A_{n-3} & C_{n-3} \\ 0 & \cdots & \ddots & \ddots & \ddots & \ddots & \ldots & d_{n-2} & a_{n-2} & A_{n-2} \\ A_{n-1} & 0 & \cdots & \ddots & \ddots & \ddots & \ldots & b_{n-1} & d_{n-1} & a_{n-1} \\ a_n & A_n & 0 & \cdots & \cdots & & D_n & B_n & b_n & d_n \end{bmatrix} \tag{1.1}$$

and

[1] Home address: Mathematics Department, Faculty of Science, Mansoura University, Mansoura, 35516, Egypt. E-mail:abibka@mans.edu.eg

$$
H^{anti} = \begin{bmatrix}
b_1 & B_1 & 0 & \cdots & & 0 & C_1 & A_1 & a_1 & d_1 \\
B_2 & 0 & 0 & \cdots & \iddots & & A_2 & a_2 & d_2 & b_2 \\
0 & \cdots & \cdots & \iddots & \iddots & & a_3 & d_3 & b_3 & B_3 \\
0 & \cdots & \iddots & \iddots & \iddots & & d_4 & b_4 & B_4 & 0 \\
\vdots & \iddots & \iddots & \iddots & \iddots & \iddots & \iddots & \iddots & \iddots & \vdots \\
0 & \iddots & \iddots & \iddots & \iddots & \iddots & \iddots & \iddots & \cdots & \vdots \\
C_{n-3} & A_{n-3} & a_{n-3} & d_{n-3} & \iddots & \iddots & \iddots & \cdots & \cdots & \\
A_{n-2} & a_{n-2} & d_{n-2} & b_{n-2} & \iddots & \iddots & \cdots & \cdots & \cdots & \\
a_{n-1} & d_{n-1} & b_{n-1} & B_{n-1} & \iddots & \cdots & \cdots & \cdots & & A_{n-1} \\
d_n & b_n & B_n & 0 & \cdots & \cdots & \cdots & & A_n & a_n
\end{bmatrix} \qquad (1.2)
$$

where $n \geq 8$.

The inverses of cyclic heptadiagonal matrices are usually required in science and engineering applications, for more details (see special cases, [1-9]). The motivation of the current paper is to establish efficient algorithms for computing determinant and inverting cyclic heptadiagonal and anti-cyclic heptadiagonal matrices of the form (1.1) and (1.2).

The paper is organized as follows. In Section 2, new symbolic computational algorithms, that will not break, is constructed. In Section 3, an illustrative example is given. Conclusions of the work are given in Section 4.

## 2. Main results

In this section we shall focus on the construction of new symbolic computational algorithms for computing the determinant and the inverse of general cyclic heptadiagonal matrices. The solution of cyclic heptadiagonal linear systems of the form (1.2) will be taken into account.

Throughout this section, the parameter t is just a symbolic name and detH is the determinant of the heptadiagonal matrix of the form (1.1). We state the following result without proof (see [10,11]).

**Theorem 2.1.** Suppose tha:

$$u_0 = 1, \;\; u_1 = |d_1| = d_1, \tag{2.1}$$

$$u_i=\begin{vmatrix}
d_1 & a_1 & A_1 & C_1 & 0 & 0 & \cdots & \cdots & \cdots & \\
b_2 & d_2 & a_2 & A_2 & C_2 & \ddots & \ddots & \cdots & \cdots & \\
B_3 & b_3 & d_3 & a_3 & A_3 & \ddots & \ddots & \ddots & \cdots & \\
D_4 & B_4 & b_4 & d_4 & a_4 & \ddots & \ddots & \ddots & \ddots & \\
0 & \ddots & \ddots & \ddots & \ddots & \ddots & \ddots & \ddots & \ddots & \vdots \\
\vdots & \ddots & \ddots & \ddots & \ddots & \ddots & \ddots & \ddots & \ddots & \\
0 & \ddots & \ddots & \ddots & \ddots & \ddots & a_{i-3} & A_{i-3} & C_{i-3} & \\
0 & \cdots & \ddots & \ddots & \ddots & \ddots & d_{i-2} & a_{i-2} & A_{i-2} & \\
0 & \cdots & \cdots & \ddots & \ddots & \ddots & b_{i-1} & d_{i-1} & a_{i-1} & \\
0 & \cdots & \cdots & \cdots & \cdots & D_i & B_i & b_i & d_i &
\end{vmatrix}, \quad i=2,3,\ldots,n-2, \tag{2.2}$$

$$u_{n-1}=\begin{vmatrix}
d_1 & a_1 & A_1 & C_1 & 0 & 0 & \cdots & \cdots & \cdots & \\
b_2 & d_2 & a_2 & A_2 & C_2 & \ddots & \ddots & \cdots & \cdots & \\
B_3 & b_3 & d_3 & a_3 & A_3 & \ddots & \ddots & \ddots & \cdots & \\
D_4 & B_4 & b_4 & d_4 & a_4 & \ddots & \ddots & \ddots & \ddots & \\
0 & \ddots & \ddots & \ddots & \ddots & \ddots & \ddots & \ddots & \ddots & \vdots \\
\vdots & \ddots & \ddots & \ddots & \ddots & \ddots & \ddots & \ddots & \ddots & \\
0 & \ddots & \ddots & \ddots & \ddots & \ddots & a_{n-4} & A_{n-4} & C_{n-4} & \\
0 & \cdots & \ddots & \ddots & \ddots & \ddots & d_{n-3} & a_{n-3} & A_{n-3} & \\
0 & \cdots & \cdots & \ddots & \ddots & \ddots & b_{n-2} & d_{n-2} & a_{n-3} & \\
A_{n-1} & \cdots & \cdots & \cdots & \cdots & D_{n-1} & B_{n-1} & b_{n-1} & d_{n-1} &
\end{vmatrix}, \tag{2.3}$$

**And**

$$u_n=\det H=\begin{vmatrix}
d_1 & a_1 & A_1 & C_1 & 0 & 0 & \cdots & \cdots & & b_1 \\
b_2 & d_2 & a_2 & A_2 & C_2 & \ddots & \ddots & \cdots & \cdots & \\
B_3 & b_3 & d_3 & a_3 & A_3 & \ddots & \ddots & \ddots & \cdots & \\
D_4 & B_4 & b_4 & d_4 & a_4 & \ddots & \ddots & \ddots & \ddots & \\
0 & \ddots & \ddots & \ddots & \ddots & \ddots & \ddots & \ddots & \ddots & \vdots \\
\vdots & \ddots & \ddots & \ddots & \ddots & \ddots & \ddots & \ddots & \ddots & \\
0 & \ddots & \ddots & \ddots & \ddots & \ddots & & a_{n-3} & A_{n-3} & C_{n-3} \\
0 & \cdots & \ddots & \ddots & \ddots & \ddots & & d_{n-2} & a_{n-2} & A_{n-2} \\
A_{n-1} & 0 & \cdots & \ddots & \ddots & \ddots & & b_{n-1} & d_{n-1} & a_{n-1} \\
a_n & A_n & 0 & \cdots & \cdots & & D_n & B_n & b_n & d_n
\end{vmatrix} \tag{2.3}$$

Then, we have the two-term recurrence

$$u_i=\alpha_i u_{i-1}, \quad i=1,2,\ldots,n \tag{2.4}$$

where

$$\alpha_i = \begin{cases} d_1 & \text{if } i = 1 \\ d_2 - f_2 g_1 & \text{if } i = 2 \\ d_3 - e_3 z_1 - f_3 g_2 & \text{if } i = 3 \\ d_i - \dfrac{D_i}{\alpha_{i-3}} C_{i-3} - e_i z_{i-2} - f_i g_{i-1} & \text{if } i = 4(5)n-2 \\ d_{n-1} - \sum_{j=1}^{n-2} w_j k_j & \text{if } i = n-1 \\ d_n - \sum_{j=1}^{n-1} v_j h_j & \text{if } i = n, \end{cases} \tag{2.5}$$

$$k_i = \begin{cases} \dfrac{A_{n-1}}{\alpha_1} & \text{if } i = 1 \\ -\dfrac{k_1 g_1}{\alpha_2} & \text{if } i = 2 \\ -\dfrac{(k_1 z_1 + k_2 g_2)}{\alpha_3} & \text{if } i = 3 \\ -\dfrac{(k_{i-3} C_{i-3} + k_{i-2} z_{i-2} + k_{i-1} g_{i-1})}{\alpha_i} & \text{if } i = 4(5)n-5 \\ \dfrac{(D_{n-1} - k_{n-7} C_{n-7} - k_{n-6} z_{n-6} - k_{n-5} g_{n-5})}{\alpha_{n-4}} & \text{if } i = n-4 \\ \dfrac{(B_{n-1} - k_{n-6} C_{n-6} - k_{n-5} z_{n-5} - k_{n-4} g_{n-4})}{\alpha_{n-3}} & \text{if } i = n-3 \\ \dfrac{(b_{n-1} - k_{n-5} C_{n-5} - k_{n-4} z_{n-4} - k_{n-3} g_{n-3})}{\alpha_{n-2}} & \text{if } i = n-2, \end{cases} \tag{2.6}$$

$$h_i = \begin{cases} \dfrac{a_n}{\alpha_1} & \text{if } i = 1 \\ \dfrac{A_n - h_1 g_1}{\alpha_2} & \text{if } i = 2 \\ -\dfrac{(h_1 z_1 + h_2 g_2)}{\alpha_3} & \text{if } i = 3 \\ -\dfrac{(h_{i-3} C_{i-3} + h_{i-2} z_{i-2} + h_{i-1} g_{i-1})}{\alpha_i} & \text{if } i = 4(5)n-4 \\ \dfrac{(D_n - h_{n-6} C_{n-6} - h_{n-5} z_{n-5} - h_{n-4} g_{n-4})}{\alpha_{n-3}} & \text{if } i = n-3 \\ \dfrac{(B_n - h_{n-5} C_{n-5} - h_{n-4} z_{n-4} - h_{n-3} g_{n-3})}{\alpha_{n-2}} & \text{if } i = n-2 \\ \dfrac{(b_n - \sum_{j=1}^{n-2} h_j w_j)}{\alpha_{n-1}} & \text{if } i = n-1, \end{cases} \tag{2.7}$$

$$v_i = \begin{cases} b_1 & \text{if } i = 1 \\ B_2 - f_2 v_1 & \text{if } i = 2 \\ -e_3 v_1 - f_3 v_2 & \text{if } i = 3 \\ -\dfrac{D_i}{\alpha_{i-3}} v_{i-3} - e_i v_{i-2} - f_i v_{i-1} & \text{if } i = 4(5)n-4 \\ C_{n-3} - \dfrac{D_{n-3}}{\alpha_{n-6}} v_{n-6} - e_{n-3} v_{n-5} - f_{n-3} v_{n-4} & \text{if } i = n-3 \\ A_{n-2} - \dfrac{D_{n-2}}{\alpha_{n-5}} v_{n-5} - e_{n-2} v_{n-4} - f_{n-2} v_{n-3} & \text{if } i = n-2 \\ a_{n-1} - \sum_{j=1}^{n-2} v_j k_j & \text{if } i = n-1, \end{cases} \tag{2.8}$$

$$w_i = \begin{cases} B_1 & \text{if } i = 1 \\ -f_2 w_1 & \text{if } i = 2 \\ -f_3 w_2 - e_3 w_1 & \text{if } i = 3 \\ -\dfrac{D_i}{\alpha_{i-3}} w_{i-3} - e_i w_{i-2} - f_i w_{i-1} & \text{if } i = 4(5)n-5 \\ C_{n-4} - \dfrac{D_{n-4}}{\alpha_{n-7}} W_{n-7} - e_{n-4} w_{n-6} - f_{n-4} w_{n-5} & \text{if } i = n-4 \\ A_{n-3} - \dfrac{D_{n-3}}{\alpha_{n-6}} w_{n-6} - e_{n-3} w_{n-5} - f_{n-3} w_{n-4} & \text{if } i = n-3 \\ a_{n-2} - \dfrac{D_{n-2}}{\alpha_{n-5}} w_{n-5} - e_{n-2} w_{n-4} - f_{n-2} w_{n-3} & \text{if } i = n-2, \end{cases} \tag{2.9}$$

$$f_i = \begin{cases} \dfrac{b_2}{\alpha_1} & \text{if } i = 2 \\ \dfrac{b_3 - e_3 g_1}{\alpha_2} & \text{if } i = 3 \\ \dfrac{b_i - \dfrac{D_i}{\alpha_{i-3}} z_{i-3} - e_i g_{i-2}}{\alpha_{i-1}} & \text{if } i = 4(5)n-2, \end{cases} \tag{2.10}$$

$$e_i = \begin{cases} \dfrac{B_3}{\alpha_1} & \text{if } i = 3 \\ \dfrac{B_i - \dfrac{D_i}{\alpha_{i-3}} g_{i-3}}{\alpha_{i-2}} & \text{if } i = 4(5)n-2, \end{cases} \tag{2.11}$$

$$g_i = \begin{cases} a_1 & \text{if } i = 1 \\ a_2 - f_2 z_1 & \text{if } i = 2 \\ a_i - f_i z_{i-1} - e_i C_{i-2} & \text{if } i = 3(4)n-3, \end{cases} \tag{2.12}$$

$$z_i = \begin{cases} A_1 & \text{if } i = 1 \\ A_i - f_i C_{i-1} & \text{if } i = 2(3)n-4. \end{cases} \tag{2.13}$$

At this point it is convenient to formulate our first result. It is a symbolic algorithm for computing the determinant of a cyclic heptadiagonal matrix H of the form (1.1)

**Algorithm 2.1.** To compute det H for the cyclic heptadiagonal matrix H in (1.1), we may proceed as follows:

**Step 1:** Set $\alpha_1=d_1$. If $\alpha_1=0$ then $\alpha_1=t$ end if. Set $u_1=d_1$, $g_1=a_1$, $z_1=A_1$, $k_1=A_{n-1}/\alpha_1$, $v_1=b_1$, $w_1=B1$, $h_1=a_n/\alpha_1$, $w_1=B_1$ , $f_2=b_2/\alpha_1$, $e_3=B_3/\alpha_1$, $\alpha_2=d_2-f_2*g_1$. If $\alpha_2=0$ then $\alpha_2=t$ end if. Set $K_2=-k_1*g_1/\alpha_2$, $u_2=\alpha_2*u_1$, $v_2=B_2-f_2*v_1$, $w_2=-f_2*w_1$, $h_2=(A_n-h_1*g_1)/\alpha_2$, $g_2=a_2-f_2*z_1$, $f_3=(b_3-e_3*g_1)/\alpha_2$, $\alpha_3=d_3-e_3*z_1-f_3*g_2$. If $\alpha_3=0$ then $\alpha_3=t$ end if. Set $u_3=\alpha_3*u_2$, $k3=-(k_1*z_1+k_2*g_2)/\alpha_3$, $h_3=-(h_1z_1+h_2g_2)/\alpha_3$, $v_3=-e_3*v_1-f_3*v_2$, $w_3=-f_3w_2-e_3w_1$.

**Step 2:** Compute and simplify:

For i from 4 to n-2 do

- $e_i=(B_i-D_i*g_{i-3}/\alpha_{i-3})/\alpha_{i-2}$
- $f_i=(bi-D_i*z_{i-3}/\alpha_{i-3}-e_ig_{i-2})/\alpha_{i-1}$
- $z_{i-2}=A_{i-2}-f_{i-2}*C_{i-3}$
- $g_{i-1}=a_{i-1}-f_{i-1}*z_{i-2}-e_{i-1}*C_{i-3}$
- $\alpha_i=(d_i-D_i*C_{i-3}/\alpha_{i-3}-e_ig_{i-2}-f_ig_{i-1})$
- If $\alpha_i=0$ then $\alpha_i=t$ end if
- $u_i=\alpha_i*u_{i-1}$

End do

**Step 3:** Compute and simplify:

For i from 4 to n-5 do

- $K_i=-(k_{i-3}*C_{i-3}+k_{i-2}*z_{i-2}+k_{i-1}*g_{i-1})/\alpha_i$
- $w_i=-(D_i*w_{i-3}/\alpha_{i-3}+e_i*w_{i-2}+f_i*w_{i-1})$

End do

**Step 4:** Compute and simplify:

For i from 4 to n-4 do

- $h_i=-(h_{i-3}*C_{i-3}+h_{i-2}*z_{i-2}+h_{i-1}*g_{i-1})/\alpha_i$
- $v_i=-(D_i*v_{i-3}/\alpha_{i-3}+e_i*v_{i-2}+f_i*v_{i-1})$

End do

**Step 5:** Compute simplify:

$k_{n-4}=(D_{n-1}-k_{n-5}*g_{n-5}-k_{n-6}*z_{n-6}-k_{n-7}*C_{n-7})/\alpha_{n-4}$

$k_{n-3}=(B_{n-1}-k_{n-4}*g_{n-4}-k_{n-5}*z_{n-5}-k_{n-6}*C_{n-6})/\alpha_{n-3}$

$k_{n-2}=(b_{n-1}-k_{n-3}*g_{n-3}-k_{n-4}*z_{n-4}-k_{n-5}*C_{n-5})/\alpha_{n-2}$

$w_{n-4}=C_{n-4}-D_{n-4}*w_{n-7}/\alpha_{n-7}-e_{n-4}w_{n-6}-f_{n-4}w_{n-5}$

$w_{n-3}=A_{n-3}-D_{n-3}*w_{n-6}/\alpha_{n-6}-e_{n-3}w_{n-5}-f_{n-3}w_{n-4}$

$w_{n-2}=a_{n-2}-D_{n-2}*w_{n-5}/\alpha_{n-5}-e_{n-2}w_{n-4}-f_{n-2}w_{n-3}$

$h_{n-3}=(D_n-h_{n-4}*g_{n-4}-h_{n-5}*z_{n-5}-h_{n-6}*C_{n-6})/\alpha_{n-3}$

$h_{n-2}=(B_n-h_{n-3}*g_{n-3}-h_{n-4}*z_{n-4}-h_{n-5}*C_{n-5})/\alpha_{n-2}$

$v_{n-3}=C_{n-3}-D_{n-3}*v_{n-6}/\alpha_{n-6}-e_{n-3}v_{n-5}-f_{n-3}v_{n-4}$

$v_{n-2}=A_{n-2}-D_{n-2}*v_{n-5}/\alpha_{n-5}-e_{n-2}v_{n-4}-f_{n-2}v_{n-3}$

$v_{n-1}=a_{n-1}-\sum_{j=1}^{n-2} k_j v_j$

$\alpha_{n-1}=d_{n-1}-\sum_{j=1}^{n-2} k_j w_j$

If $\alpha_{n-1}=0$ then $\alpha_{n-1}= t$ end if

$u_{n-1}= \alpha_{n-1}*u_{n-2}$

$h_{n-1}=(b_n-\sum_{j=1}^{n-2} h_j w_j)/ \alpha_{n-1}$

$\alpha_n=d_n-\sum_{j=1}^{n-1} h_j v_j$

If $\alpha_n=0$ then $\alpha_n= t$ end if

$u_n= \alpha_n*u_{n-1}$

**Step 6:** Compute det H $=(u_n)_{t=0}$.

Now we formulate a second result. Following [10], we can using **CHINV** algorithm to compute the inverse of a general cyclic heptadiagonal matrix of the form (1.1) when it exists. The following theorem given the inverse of cyclic anti-heptadiagonal matrix $H^{anti}$.

**Theorem 2.2.** Let P be an n X n matrix as the following form

$$P=\begin{bmatrix} 0 & \cdots & \cdots & & 1 \\ 0 & & \iddots & & 0 \\ \vdots & \iddots & \iddots & \iddots & \vdots \\ 0 & 1 & \iddots & & 0 \\ 1 & 0 & \cdots & \cdots & \end{bmatrix}$$

Then the inverse of cyclic anti-heptadiagonal matrix $H^{anti}$ is given by

$$H^{anti^{-1}} = PH^{-1} \tag{2.14}$$

## 3. An illustrative example

In this section we give an example for the sake of illustration.

**Example 3.1.** Consider the 10 X 10 cyclic heptadiagonal and cyclic anti- heptadiagonal matrices

$$A=\begin{bmatrix} 1 & -1 & 1 & -2 & 0 & 0 & 0 & 0 & 5 & -4 \\ 2 & 1 & 4 & 1 & -5 & 0 & 0 & 0 & 0 & 1 \\ 2 & 1 & -1 & 1 & 2 & 3 & 0 & 0 & 0 & 0 \\ 2 & -2 & 3 & 1 & 5 & -6 & 0 & 0 & 0 & 0 \\ 0 & 1 & 1 & 7 & 1 & 8 & 1 & 2 & 0 & 0 \\ 0 & 0 & -1 & -1 & -9 & -1 & -1 & -1 & 1 & 0 \\ 0 & 0 & 0 & 2 & 2 & 6 & 2 & 3 & 1 & -3 \\ 0 & 0 & 0 & 0 & -2 & -2 & 1 & 1 & 3 & 5 \\ 6 & 0 & 0 & 0 & 0 & 3 & 1 & 3 & 4 & -1 \\ 1 & 4 & 0 & 0 & 0 & 0 & 2 & 3 & 4 & 1 \end{bmatrix}, \tag{3.1}$$

and

$$B=\begin{bmatrix} -4 & 5 & 0 & 0 & 0 & 0 & -2 & 1 & -1 & 1 \\ 1 & 0 & 0 & 0 & 0 & -5 & 1 & 4 & 1 & 2 \\ 0 & 0 & 0 & 0 & 3 & 2 & 1 & -1 & 1 & 2 \\ 0 & 0 & 0 & 0 & -6 & 5 & 1 & 3 & -2 & 2 \\ 0 & 0 & 2 & 1 & 8 & 1 & 7 & 1 & 1 & 0 \\ 0 & 1 & -1 & -1 & -1 & -9 & -1 & -1 & 0 & 0 \\ -3 & 1 & 3 & 2 & 6 & 2 & 2 & 0 & 0 & 0 \\ 5 & 3 & 1 & 1 & -2 & -2 & 0 & 0 & 0 & 0 \\ -1 & 4 & 3 & 1 & 3 & 0 & 0 & 0 & 0 & 6 \\ 1 & 4 & 3 & 2 & 0 & 0 & 0 & 0 & 4 & 1 \end{bmatrix} \tag{3.2}$$

By using algorithm 2.1 and theorem 2.1 we obtained the determinant of matrix (3.1) and inverse of matrix (3.2) respectively:

$$\det A=\left.\frac{9(164542t+298485)(-13214-2292t)}{(6607+1146t)}\right|_{t=0}=-2686365,$$

$$B^{-1}=RA^{-1}=\begin{bmatrix} -\frac{629}{19899} & \frac{6559}{298485} & -\frac{524}{5427} & -\frac{33313}{298485} & \frac{10348}{298485} & -\frac{8794}{59697} & -\frac{39851}{298485} & \frac{34688}{298485} & \frac{46}{603} & -\frac{16037}{298485} \\ \frac{3269}{19899} & -\frac{16393}{298485} & -\frac{76}{5427} & -\frac{6254}{298485} & \frac{26744}{298485} & -\frac{875}{59697} & -\frac{33343}{298485} & \frac{20884}{298485} & -\frac{1}{603} & \frac{7589}{298485} \\ -\frac{304}{1809} & -\frac{1076}{5427} & -\frac{6476}{5427} & -\frac{1549}{5427} & \frac{1606}{5427} & -\frac{1781}{5427} & -\frac{2873}{5427} & -\frac{1714}{5427} & \frac{115}{201} & \frac{484}{5427} \\ -\frac{478}{19899} & \frac{88904}{298485} & \frac{8642}{5427} & \frac{164542}{298485} & -\frac{190852}{298485} & \frac{35182}{59697} & \frac{384389}{298485} & \frac{161773}{298485} & -\frac{484}{603} & -\frac{12862}{298485} \\ \frac{287}{6633} & \frac{5713}{99495} & -\frac{50}{1809} & -\frac{15106}{99495} & \frac{2086}{99495} & -\frac{2665}{19899} & -\frac{3587}{99495} & \frac{2621}{99495} & \frac{28}{603} & -\frac{7739}{99495} \\ \frac{559}{19899} & -\frac{10469}{298485} & -\frac{191}{5427} & -\frac{7672}{298485} & \frac{8467}{298485} & -\frac{7507}{59697} & -\frac{25694}{298485} & -\frac{5308}{298485} & \frac{16}{603} & \frac{1387}{298485} \\ -\frac{226}{19899} & -\frac{23047}{298485} & \frac{965}{5427} & \frac{60034}{298485} & \frac{33266}{298485} & \frac{12943}{59697} & \frac{16748}{298485} & \frac{2941}{298485} & -\frac{64}{603} & \frac{13346}{298485} \\ \frac{475}{6633} & \frac{18004}{99495} & -\frac{446}{1809} & -\frac{10423}{99495} & \frac{4513}{99495} & -\frac{4516}{19899} & -\frac{11486}{99495} & -\frac{3217}{99495} & \frac{10}{201} & -\frac{2927}{99495} \\ -\frac{124}{19899} & \frac{2026}{59697} & \frac{284}{5427} & -\frac{2581}{59697} & \frac{1549}{59697} & -\frac{1586}{59697} & -\frac{7871}{59697} & -\frac{8353}{59697} & -\frac{28}{603} & \frac{11866}{59697} \\ -\frac{107}{2211} & \frac{45}{737} & \frac{68}{201} & \frac{90}{737} & -\frac{78}{737} & \frac{87}{737} & \frac{266}{2211} & \frac{20}{737} & \frac{1}{201} & -\frac{18}{737} \end{bmatrix}$$

## 4. Conclusions

In this work new computational algorithms have been developed for computing the determinant and inverse of general cyclic anti-heptadiagonal matrices. The algorithms are reliable, computationally efficient and will not fail. The algorithms are natural generalizations of some algorithms in current use.